\documentclass[5p,times]{elsarticle}

\usepackage{listings}
\usepackage{xcolor}

\usepackage[noend]{algorithm2e}

\usepackage{siunitx}

\usepackage{amsmath}

\usepackage{booktabs} 
\usepackage{tabularx}

\usepackage[hyphens]{url}
\usepackage{hyperref}
\hypersetup{hidelinks, 
  colorlinks=true, 
  pdfauthor={Adrian Nilsson, Simon Smith, Jonas Hagmar, Magnus Önnheim, Mats Jirstrand},
  }

\usepackage{float}

\definecolor{codegreen}{rgb}{0,0.6,0}
\definecolor{codegray}{rgb}{0.5,0.5,0.5}
\definecolor{codepurple}{rgb}{0.58,0,0.82}
\definecolor{backcolour}{rgb}{0.95,0.95,0.92}
\lstdefinestyle{mystyle}{
  backgroundcolor=\color{backcolour},   commentstyle=\color{codegreen},
  keywordstyle=\color{magenta},
  numberstyle=\tiny\color{codegray},
  stringstyle=\color{codepurple},
  basicstyle=\ttfamily\footnotesize,
  breakatwhitespace=false,         
  breaklines=true,                 
  captionpos=b,                    
  keepspaces=true,                 
  numbers=left,                    
  numbersep=5pt,                  
  showspaces=false,                
  showstringspaces=false,
  showtabs=false,                  
  tabsize=2
}
\begin{document}

\begin{frontmatter}


\title{The AutoSPADA Platform: User-Friendly Edge Computing for Distributed Learning and Data Analytics in Connected Vehicles}


\author[1]{Adrian Nilsson}
\author[1]{Simon Smith}
\author[1]{Jonas Hagmar}
\author[1]{Magnus Önnheim}

\author[1,2]{Mats Jirstrand\corref{cor1}}
\ead{jirstran@chalmers.se}

\cortext[cor1]{Corresponding author}

\affiliation[1]{organization={Fraunhofer-Chalmers Centre},
            city={Gothenburg},
            postcode={SE-412 88}, 
            country={Sweden}}
\affiliation[2]{organization={Department of Electrical Engineering, Chalmers University of Technology},
            city={Gothenburg},
            postcode={SE-412 96}, 
            country={Sweden}}

\begin{abstract}

Contemporary connected vehicles host numerous applications, such as diagnostics and navigation, and new software is continuously being developed.
However, the development process typically requires offline batch processing of large data volumes.
In an edge computing approach, data analysts and developers can instead process sensor data directly on computational resources inside vehicles. 
This enables rapid prototyping to shorten development cycles and reduce the time to create new business values or insights.
This paper presents the design, implementation, and operation of the AutoSPADA edge computing platform for distributed data analytics.
The platform's design follows scalability, reliability, resource efficiency, privacy, and security principles promoted through mature and industrially proven technologies.
In AutoSPADA, computational tasks are general Python scripts, and we provide a library to, for example, read signals from the vehicle and publish results to the cloud.
Hence, users only need Python knowledge to use the platform.
Moreover, the platform is designed to be extended to support additional programming languages.

\end{abstract}




\end{frontmatter}



\section{Introduction}

Today, network-connected devices outnumber the global population threefold~\cite{cisco2020:forecast}.
Around half of these are Internet of Things (IoT) devices such as connected vehicles, home automation systems, and industrial smart sensors.
Moreover, connected vehicles are forecasted to be the fastest growing IoT application with $30\%$ yearly growth between 2018 and 2023~\cite{cisco2020:forecast}.
This growth further exacerbates the problems faced in big automotive data~\cite{johanson2014:bigautomotivedata}.
For example, the per year data collection need of Volvo Car Corporation and Volvo Group Trucks Technology is already estimated to be in the order of exabytes~\cite{Petersson2020:oodida}.
As the data volume and velocity continue to increase, offline batch processing becomes increasingly challenging, inefficient, or even infeasible.
New methods and workflows are, therefore, needed to transform the rich edge-generated data into insights and business value.

Edge computing is a paradigm that emerged in response to the surge in the number of IoT devices and the massive amount of data they produce.
In edge computing, data is processed close to the source of the data rather than sent to the cloud for processing~\cite{shi2016:edgecomputing}, making it naturally positioned to protect the privacy of data owners.
This paradigm can reduce data transfer costs, lower energy consumption, and improve latency~\cite{fang2019:edgesurvey, yuyi2017:surveyMEC} and has led to the development of novel privacy-preserving algorithms and analytics as well as new applications of artificial intelligence~\cite{elkordy2023:faSurvey, nguyen2021:FL-IoT, deng2020:confluenceECAI, zhi2019:lastmileEI}.
However, distributed edge computing applications face several system-level challenges such as scalability, device and data heterogeneity, and device reliability.
For example, applications must continue to function even if the network is unavailable and the cloud must be able to orchestrate the possibly large number of devices.
Therefore, edge computing applications usually rely on a framework or platform with components deployed both to the cloud and the edge.

This paper presents the AutoSPADA edge computing platform for distributed data analytics.
The platform aims to seamlessly connect data analysts with the data and computational resources of edge devices such as those found in contemporary vehicles.
Users do not have to be Kubernetes experts or proficient in low-level or specialized programming languages.
Instead, AutoSPADA users write ordinary Python programs, significantly lowering the barrier of entry.
A data scientist, researcher, or engineer may already have Python scripts for offline data analysis that could be transformed into a program ready to be deployed to AutoSPADA edge clients. 
We provide users with a Python library to, among other things, read signal values directly from the edge device and publish results to the cloud.

AutoSPADA is designed to support a wide range of paradigms including data aggregation, monitoring, and machine learning.
Results from clients are collected by the server infrastructure and can be received interactively through streaming channels or retrieved on demand.
An important concern has been to enforce privacy and security throughout the platform to prevent attacks by a third party, or even from malicious tasks.
Also, the architecture is designed to be scalable since the platform is intended to be deployable to many clients.

\section{Background}
The AutoSPADA (Automotive Stream Processing and Distributed Analytics) project~\cite{grimmemyhr2023:autospada} is a continuation of the OODIDA (On-board/Off-board Distributed Data Analytics) project~\cite{Petersson2020:oodida, Ulm2021:OODIDA}. 
Although conceptually similar, the design of the AutoSPADA platform developed in the project is in many ways different from that of OODIDA.
For AutoSPADA, we wanted to create an edge computing platform characterized by a high degree of interactivity, flexibility, and ease of use.
In particular, users should not be required to learn any new software tools, paradigms, or languages---the platform should be accessible to anyone comfortable writing Python code.
Therefore, we have focused on enabling users to deploy general Python scripts directly to hardware at the edge.
In contrast, OODIDA only had experimental support for deploying code to edge clients~\cite{ULM2020:rapidprototyping}, otherwise limiting users to a set of fixed-function aggregations.

A stated goal of AutoSPADA is to elevate the platform from a prototype to a pre-production system designed and ready for deployment at scale.
Taking this step places additional demands on the platform's overall design and supporting technologies.
The remainder of this section discusses the limitations of the design and technological choices made in OODIDA and argues that a revised design is required to reach a higher level of technological maturity.

\subsection{Implementation language}

OODIDA is a distributed application written in Erlang with an additional Python component on edge devices. 
In an Erlang application, processes can be executed seamlessly on any computational node.
This facilitates the rapid development of distributed applications since the communication between processes is abstracted by the language.
However, this choice of languages is not ideal for realizing the design goals of the platform, which is also discussed to some degree in the OODIDA retrospective and future vision~\cite{Ulm2021:OODIDA}.

Using Erlang and Python in the OODIDA implementation has several limitations.
One is the need for separate runtimes, specifically, the BEAM virtual machine and the Python interpreter.
These runtimes offer portability but consume additional disk space and memory, which is especially undesirable in a constrained edge environment.
Another limitation is that both Erlang and Python are dynamically typed languages.
Using statically typed languages eliminates large classes of type errors at compile time that dynamically typed languages fail to detect.
For the AutoSPADA platform, we prioritize the type safety and performance of a compiled and statically typed language over the convenience and portability of Erlang and Python.

Maintainability is a major consideration in the design of the AutoSPADA platform. 
Using a language with a low barrier of entry and easy access to a rich ecosystem of official third-party libraries is an important maintainability aspect.
Erlang has excellent support for concurrency, but it is not a very popular language~\cite{cass2022:toplanguages} and does not have the best ecosystem of third-party packages.
For example, Erlang lacks official support for central libraries such as Docker Engine, MongoDB, and Protocol Buffers. 
Therefore, keeping Erlang as our implementation language would have forced the application to rely on possibly poorly maintained unofficial packages.
Also, using languages with widespread use has many advantages, such as making it easier to hire developers in a market where software development skills are already in high demand.

\subsection{Communication}

OODIDA relies on the built-in network capabilities of the Erlang language.
In contrast, AutoSPADA uses a proven IoT-oriented protocol with a minimal network footprint for notifications.
A binary remote procedure call (RPC) framework is used for updates and data transfers, where server and client APIs are compiled from a shared interface specification to minimize protocol mismatches between communicating parties.
For communication between components written in different languages, OODIDA instead relies on JSON messages.
Being a text-based format, JSON interfaces are inefficient and have a high development and maintenance overhead.

\subsection{Resiliency}

OODIDA is built on direct communication with clients, assuming that these are available during the task lifetime.
No such assumption is made in the AutoSPADA platform.
Instead, the application state is centralized, and state updates from clients are cached so that intermittent availability or poor network conditions do not jeopardize the integrity of the platform.

\subsection{Privacy and security}

AutoSPADA has a strong focus on privacy and security.
All network communication is secured by state-of-the-art encryption, communicating parties are always mutually authenticated, and tasks are isolated from their hosts through containerization.
OODIDA did not focus on these aspects to the same degree.
For example, Erlang was originally designed to run on private networks~\cite{armstrong2007:erlanghistory}.
Although modern-day Erlang supports communication over the Transport Layer Security (TLS) protocol, this heritage makes it difficult to use distributed Erlang securely~\cite{rodrigues2018:towards}.
Being a prototype system, OODIDA does not authenticate clients or users and lacks a user privilege system.
Moreover, OODIDA does not containerize the tasks that run on the client, which increases the risk that tasks destabilize the client by, e.g., excessive consumption of host resources.

\section{Platform Design}
\begin{figure}[ht]
  \centering
  \includegraphics[width=0.85\linewidth]{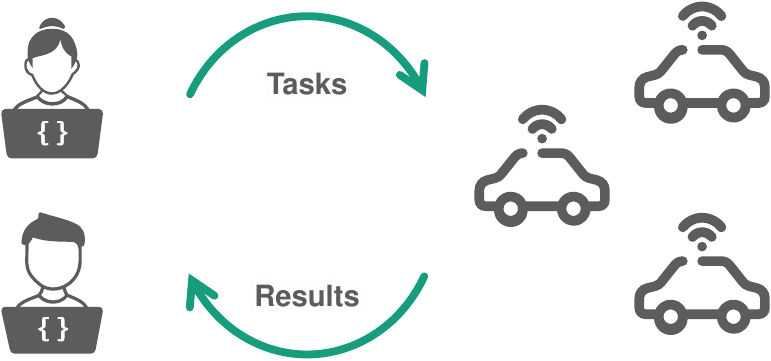}
  \caption{
The logical AutoSPADA actors are the users (left) and the clients (right).
The server nodes implement the infrastructure for transmitting tasks from users to clients and results from clients back to users.
}
  \label{fig:actors}
\end{figure}

The AutoSPADA platform is built around three distributed nodes: users, servers, and clients.
Client nodes are responsible for spawning tasks on demand and reporting the results of these tasks.
The user nodes submit tasks to run on the clients and retrieve task results from the server, streaming or on demand.
The server nodes bridge the gap between the client and user nodes by receiving and persisting task requests from users and results from clients.
This organization of nodes mirrors the actor roles in the platform, illustrated in Figure~\ref{fig:actors}.

The server and client nodes are implemented using the Go programming language.
A contributing factor to choosing Go for the core parts of the AutoSPADA platform is its widespread use and simple structure, which lowers the threshold for new developers.
The language also offers a large ecosystem of third-party libraries with official support for required APIs to reduce development time.

\subsection{Design concerns and architectural choices}

To reach a higher level of technological maturity, the AutoSPADA platform needs an underlying architecture designed with production-scale use in mind.
This requires the design to be founded on solid principles of scalability, reliability, resource efficiency, privacy, and security.
We will now discuss each of these aspects and motivate key technological choices made to address them.

\subsection{Scalability}

The AutoSPADA platform should be able to have many client nodes, which makes scalability an important factor in the architecture's design.
The scalability of distributed systems is often understood as sustaining quality of service when faced with increased workloads by adding more system resources~\cite{Lehrig2015:scalability}.
Two ways of increasing a system's resources are commonly referred to as \emph{vertical} and \emph{horizontal} scaling.
If the resources are servers in a cluster, vertical scaling means upgrading the existing servers with more powerful hardware, while horizontal scaling means adding more servers.

Horizontal scalability is usually preferred over vertical scalability.
One reason is that there is a limit to how much vertical scaling one can achieve---a motherboard can only fit so much RAM and have so many processors.
In contrast, horizontal scaling is facilitated by cloud providers that give access to practically unlimited numbers of virtual machine instances.
Moreover, horizontal scaling does not require downtime and increases redundancy for better fault tolerance.
However, horizontal scaling introduces complexity since the incoming workload must be load-balanced among the available resources to be effective.
In practice, load-balancing is achieved through the use of an orchestrator such as Kubernetes, which by itself is a non-trivial tool to master.

To achieve the high degree of scalability the platform needs, the server nodes must adhere to the stateless paradigm.
The stateless property of the server instances means that they do not retain any application state between requests.
Instead, the necessary state is read from a common database for each request, and any modified state is written to the database before the end of the request.
The benefit of stateless servers is that they become ephemeral, i.e., they only need to live for the duration of a single request.
This means that servers are horizontally scalable and that malfunctioning or superfluous instances can be taken down with little prior notice.
The pool of server instances can therefore be quickly scaled up or down to meet surges in request volume while holding the spare capacity, and thereby the cost, to a minimum.

\subsubsection{Scalable databases}
Having many stateless server nodes is only useful if the database can support them, which means that having a scalable database is of equal importance.
Specifically, a scalable database should at least have a shared-nothing architecture with automatic sharding and shard replication~\cite{stonebraker2011:10rules}.
Traditional relational database management systems such as MySQL and PostgreSQL typically share primary memory and disk storage, which makes them unsuitable for the AutoSPADA platform.

Although scalable relational databases exist~\cite{cattell2011:saclable}, e.g. MySQL Cluster, Citrix, and the Citus PostgreSQL extension~\cite{cubukcu2021:citus}, so-called NoSQL databases are often designed to address specific challenges of scaling relational databases.
For example, Dynamo~\cite{amazon2007:dynamo} was designed to achieve higher availability than previously possible in strongly consistent shared-memory relational databases.
At the time of writing, the most popular database in the NoSQL family is MongoDB~\cite{mongodb:ranking}.
Because MongoDB has a shared-nothing architecture, support for automatic sharding with configurable sharding policies as well as automated replication, failover, and recovery~\cite{mongodb2018:architecture}, it can be used as the foundation for scalable systems~\cite{stonebraker2011:10rules}.

For AutoSPADA, we have chosen MongoDB as our database since it is a proven choice in demanding applications~\cite{khan2023:nosql} and is one of the few NoSQL databases to offer per-document, multi-document, and distributed transactions~\cite{mongodb2020:transactions}.
In particular, distributed transactions are essential to the integrity of the platform.
Moreover, MongoDB has an official Go library, and because it is widely used and open source, we limit the risk of being subject to vendor lock-in.

\subsection{Reliability}

Scalability also ties into reliability since enough healthy server instances always need to be available to handle a massive number of requests from clients and users.
The ability to quickly scale the pool of stateless server instances up or down translates to a reliable backend service since a specified level of spare capacity can be easily met.
The user, server, and client nodes also need to implement robust error handling.
Moreover, we make use of proven third-party components where possible.
Using tested and established components helps to ensure that our platform operates reliably while also making it easier to maintain.

Some of our previous technological choices also cover aspects of reliability.
With the MongoDB database, it is possible to have replicas of the partitioned dataset, allowing for redundancy and, therefore, reliable operation.
One of the central features of Go, the language chosen for the server and client nodes, is the very explicit error handling, unlike languages such as C++ where unexpected exceptions can be thrown from third-party code.
This facilitates code robustness and, consequently, reliable operation of the services.

\subsubsection{Reliable operation of the client node}
While clients are not directly callable by the server, they need to quickly respond to changes in their assigned set of tasks.
Clients achieve this by subscribing to a per-client topic on a message broker.
When the application state affecting a client has changed, the server notifies the client through the message broker that it needs to dial in to retrieve the updated state.

Client devices cannot be expected to always be online since the mobile connections used for communicating with the server are inherently unreliable.
Therefore, designing the task lifetime around remote procedure calls is not suitable for the platform.
Consequently, a state-based approach was chosen, where the state of a task, including its results, is persisted in the centralized database.
The client also persists results locally until they are confirmed to be recorded in the database.
This makes the application resilient against any disruptions in communication.

\subsection{Resource constraints}

Clients connecting to the AutoSPADA platform are expected to do so over slow and unreliable network connections.
Furthermore, with a large number of clients, only a slight increase in the per-client network requirements leads to a large increase in the total network capacity requirements, potentially risking network budget issues.
The design of the application must, therefore, take the strict network constraints into account to give control over the amount of network traffic.

While client nodes are expected to be powerful enough to run an embedded Linux OS, the processing power and storage space available in the node are nevertheless limited.
These resources should primarily be used for running tasks, which means that the client program that manages tasks has to be efficient.
Therefore, the client code should preferably be compiled, native code rather than interpreted to avoid both low performance and the bloat of a runtime system.
Also, the amount of CPU and RAM that a task can allocate needs to be controllable by the application.

The Go programming language has been shown to be an energy-efficient language, at the very least on an x86 processor~\cite{Pereira2017:energy}.
This makes Go a fitting language for writing programs that target the resource-constrained hardware typically found in edge devices.

\subsubsection{Communication protocols}

To address the resource constraints imposed by mobile connections and a possibly restricted network budget, MQTT---a communication protocol designed specifically for IoT applica\-tions---was chosen for server-to-client communication.
The MQTT protocol has low message size overhead and is widely adopted~\cite{Naik2017:mqtt}.
The protocol also has configurable quality of service (QoS) levels, providing message delivery guarantees at the expense of performance.

The MQTT protocol requires a message broker to route messages between MQTT clients.
Many message brokers support MQTT, including ActiveMQ, EMQX, Mosquitto, and RabbitMQ.
However, for reliability and scalability, we require a message broker that can be configured to run as a distributed cluster.
Moreover, because MQTT sends data as plain text, the chosen broker must support TLS.
A benchmark of popular distributed MQTT brokers found only small differences between the cluster-based brokers, including EMQX and RabbitMQ~\cite{Longo2022:mqttbrokers}.
This suggests that any of the most widely used MQTT brokers are suitable alternatives.

In AutoSPADA, we chose RabbitMQ as our message broker.
Although originally designed for the AMQP protocol, RabbitMQ also has an MQTT plugin and TLS support. 
With RabbitMQ, the more feature-rich AMQP protocol can simultaneously be used between user and server nodes.
This is facilitated by official AMQP client libraries for both Go and Python. 
RabbitMQ has a maintainability advantage since it is open-source software and combines the features of MQTT and AMQP into one technology.
The main downside to RabbitMQ's MQTT plugin is that the highest QoS level is unsupported.

The information passed to the client by the broker is kept to a minimum, only consisting of the current version number of the client state.
The database structure was designed with the immutability of, e.g., task payloads in mind.
This makes it possible for clients to cache these entries locally.
Hence, network traffic is further reduced since the client can avoid repeatedly querying the database for the same information.

Many distributed applications rely on JSON for message encoding, which is less space-efficient than binary encodings and adds significant network overhead.
A case study showed that using Protocol Buffers (protobuf) to encode over 50~000 messages from a vehicle tracking system reduced the total data amount by five times compared to BSON (Binary JSON) and nearly six times compared to JSON~\cite{Popic2016:protobuf}.
Other binary protocols, such as Apache Thrift, Apache Arvo, and Microsoft Bond, are also space efficient compared to JSON~\cite{viotti2022:benchmark}.
However, because of Protocol Buffers' maturity, widespread use in industry~\cite{viotti2022:benchmark}, and official open-source libraries for both Go and Python, we chose to use protobuf in the AutoSPADA network.

\subsection{Privacy}

Data transmitted in the AutoSPADA network is inherently sensitive.
Therefore, it is of utmost importance that communication is protected end-to-end with strong and proven encryption.
To avoid malicious actors in the network, authentication mechanisms must be in place to prove the identity of clients, users, and servers.
Ideally, the system should offer mechanisms for the anonymization of client data.
This is likely very difficult to achieve. 
As a compromise, the application should offer a privilege system where only authorized users can request client results.

\subsubsection{Authorizathion and authentication protocols}

All communication in the AutoSPADA network has been designed to be encrypted using TLS, which is the standard for secure communication on the web.
TLS further supports authentication of both the server and client using X.509 certificates, which is necessary to establish the trust needed to exchange information between these parts of the network.

OpenID Connect (OIDC) was chosen as the authentication protocol to establish user identities.
OIDC is a state-of-the-art protocol that allows the authentication process to be delegated to a third-party service.
This allows the administrators of AutoSPADA deployments to configure and customize the user authentication to their needs.
Also, the OIDC protocol supports privilege systems where user resource access can be arbitrarily controlled.

For the AutoSPADA platform, relying on standard authentication protocols and delegating the difficult task of securing and maintaining user credentials to proven third-party services translates into a higher level of privacy.

\subsection{Security}

An important security concern is the isolation of user-defined tasks from the host environment.
This isolation is important not only from the perspective of preventing malicious tasks from acquiring sensitive information but also from compromising the host system by, e.g., excessive resource allocation.
Both issues are addressed by running the tasks in containers since they provide a strong boundary between active tasks and the host system and enable per-task resource limits for RAM and CPU usage.

We use the Docker platform for containerization because of its extensive toolkit that includes networking and logging features.
These features are unavailable if we interact with the container runtime directly.
By using, Docker we limit the number of dependencies and reduce code complexity for better maintainability.

\section{Platform Architecture and Implementation}

\begin{figure*}[ht]
  \centering
  \includegraphics[width=.75\linewidth]{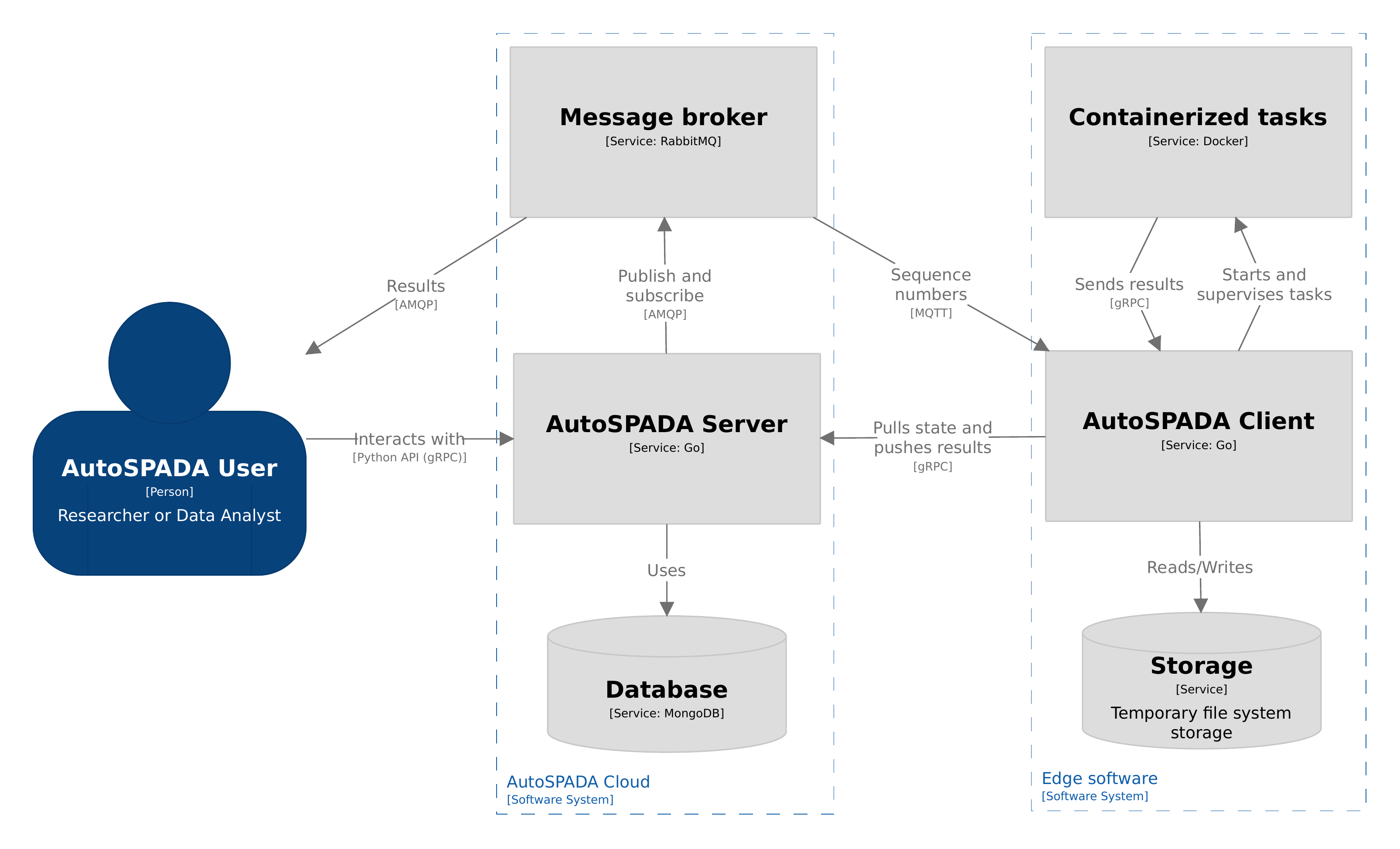}
  \caption{An overview of the AutoSPADA architecture showing the services, communication paths, languages, and protocols used in the platform.
All communication is secured by TLS.
Authentication between nodes is performed by OIDC using JSON Web Tokens (JWTs) for user nodes and through mutual TLS (mTLS) using X.509 certificates elsewhere.}
  \label{fig:structurizr-architecture}
\end{figure*}

Having discussed the various design concerns and the technology choices made to address these, we summarize the roles of the nodes of the platform and the protocols they use.
This overall architecture is illustrated in Figure~\ref{fig:structurizr-architecture}.
Users interact with server nodes through a Python library that wraps gRPC calls and provides higher-level functionality.
Users can also monitor AMQP topics to receive streaming task updates or query historical results via the Python library.
The server nodes implement gRPC services for the client and user-facing APIs and use a MongoDB database for persistent storage.
The server produces state update notifications via RabbitMQ and an MQTT bridge that clients consume.
The state update notification is a running count of the state revision for the individual client.
Through this mechanism, the client is notified about any relevant state updates,  triggering a query to the server for the latest revision of the state.
Clients cache results locally until they can be delivered to the server via gRPC.

With an understanding of the components of our architecture, the rest of this section is dedicated to its implementation.
We especially focus on the details of the client node implementation and omit the server node since it mostly implements the gRPC services.
However, we start with a brief discussion of a simplified version of our data model to better understand the resources that users create.

\subsection{Application state data model}
\begin{figure}[!ht]
    \centering
    \includegraphics[width=\linewidth]{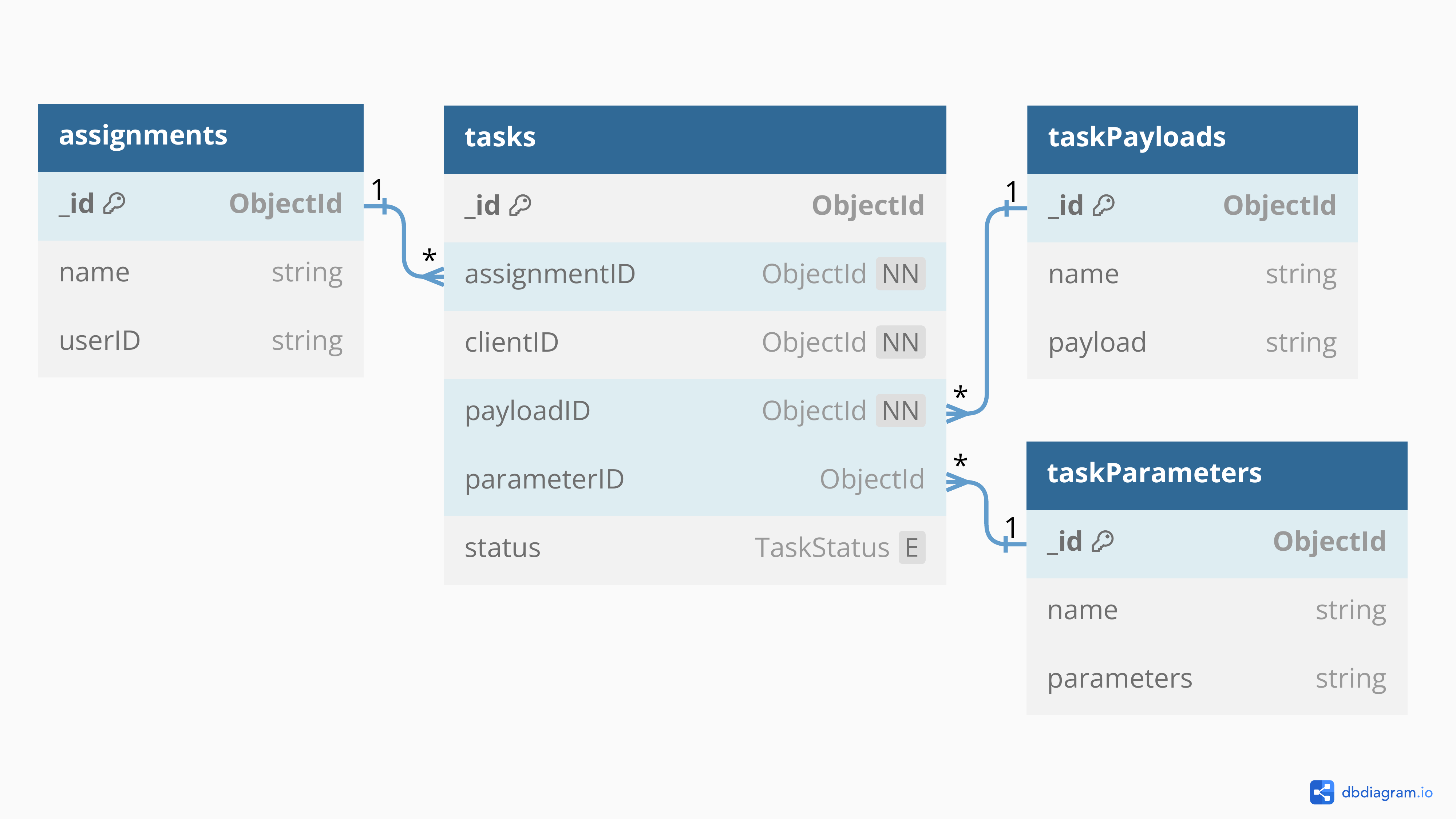}
    \caption{An entity-relationship diagram showing a simplified view of selected documents in the database, highlighting their relations. An assignment has many client-specific tasks. A task, in turn, has a payload and, optionally, a parameters document. NN is short for \textit{not null} and E is for \textit{enum}.}
    \label{fig:assignment-db}
\end{figure}

Figure~\ref{fig:assignment-db} shows a simplified view of the data model for the centralized state to help clarify our distinction between assignments, tasks, payloads, and parameters. 
Users create assignment documents containing a set of tasks.
Tasks, in turn, reference their assignment, a payload (the code to be executed), parameters, and the ID of the client for which the task is intended. 
The optional parameters document holds a JSON-serializable value that the payload can read via our client Python library.
This feature is useful to, for example, distribute a model to many clients or have the same payload use different signal names on different clients.

\subsubsection{Task life cycle}
The simplified data model in Figure~\ref{fig:assignment-db} also shows that task documents have a status field.
This field has one of four possible statuses: \lstinline{ACTIVE}, \lstinline{FINISHED},  \lstinline{ERROR}, or \lstinline{CANCELED}.
Tasks are said to be \lstinline{ACTIVE} upon creation, and the only valid transition is from \lstinline{ACTIVE} to one of the other three statuses.

Transitions from \lstinline{ACTIVE} to a \lstinline{FINISHED} or \lstinline{ERROR} status are client-initiated.
Should the payload encounter a runtime error, the task container exits with an error code and the client subsequently uploads the container logs along with an \lstinline{ERROR} status to the server.
If the task instead runs to completion without any errors, the client reports a \lstinline{FINISHED} status back to the server.
The server only accepts results from \lstinline{ACTIVE} tasks, which means that incoming results for a non-active task are ignored.

Because task payloads are general Python programs, users can define payloads that never terminate.
Such indefinite tasks only stop if explicitly instructed to do so.
Users do this by canceling the task through a Python library.
Canceling a task causes the client to stop the corresponding Docker container, forcing it to exit if needed.
Only active tasks can be canceled, however.

\subsection{Implementation of the client node}

\begin{figure*}[ht]
  \centering
  \includegraphics[width=.75\linewidth]{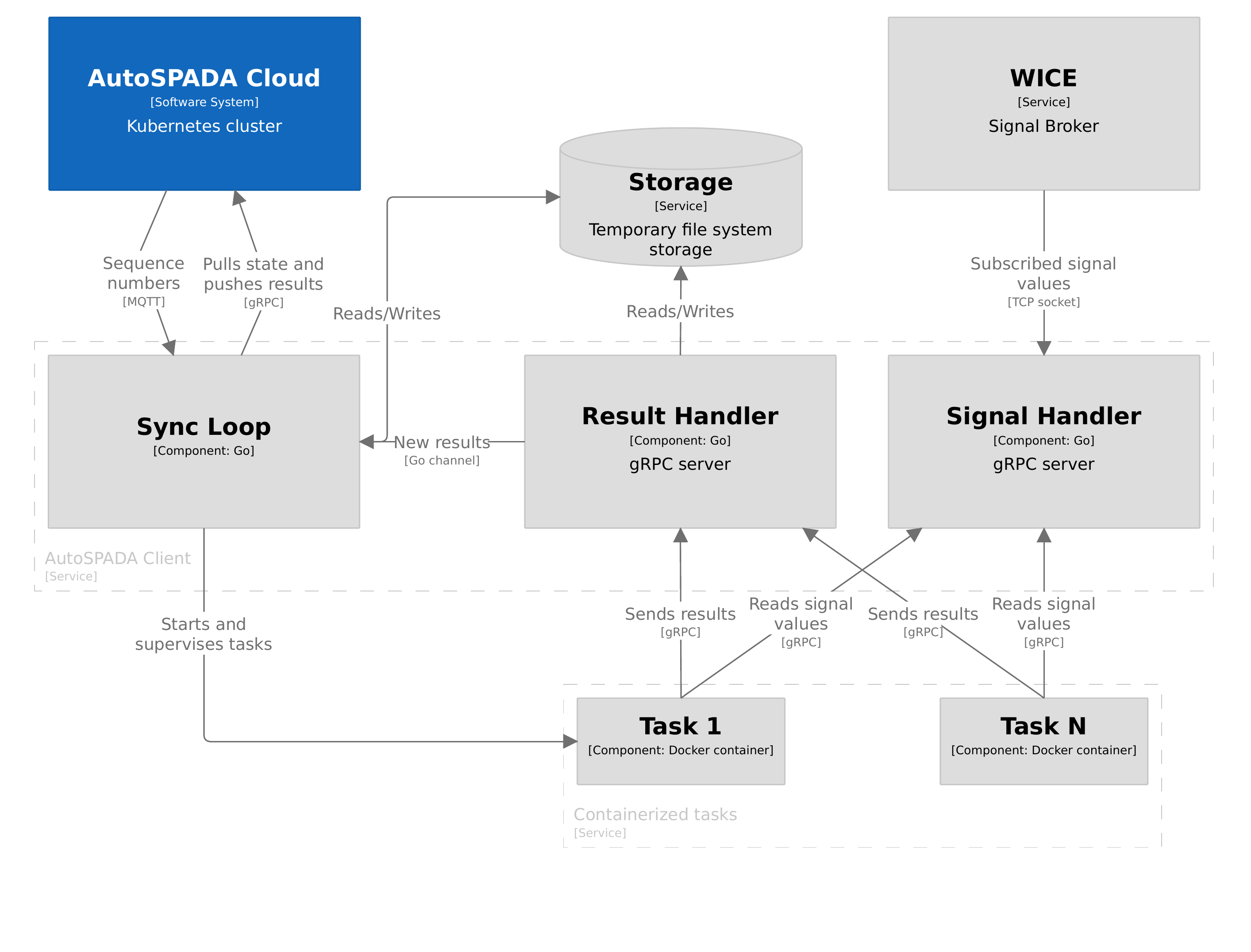}
  \caption{Detailed view of the software components in an edge device running the AutoSPADA client. The major components are the sync loop, result handler (gRPC), signal handler(gRPC), Docker, and the WICE Signal Broker. A container supervisor thread is also started for each task but is not visualized with a component box.
  }
  \label{fig:client}
\end{figure*}

The client application manages its assigned active tasks and serves the client gRPC API used by the tasks.
Figure~\ref{fig:client} gives an overview of the main components of the client.
The sync loop synchronizes the tasks' states with the centralized server state.
As part of this loop, it will spawn a thread to supervise the exit condition for each new task it starts.
The tasks can request signal values from the signal handler and send results to the result handler.
The signal handler is an abstraction layer to the actual signal source and has a state for each signal containing the latest observed value.
Keeping the latest values in memory makes it simple to determine the present value of stateful and infrequent signals, such as binary values representing an on or off state. 

\subsubsection{Synchronization Loop}
The primary responsibility of the sync loop is to keep the local state of tasks on a client synchronized with its 
centralized state stored in the server-side database in a secure, responsive, and lightweight manner.
The responsiveness and lightness come from short MQTT messages sent from the server to keep track of user-initiated changes. 
The client-local state changes when a task is created, canceled, finished, or publishes a result.
Task creation and cancelation can only be initiated by a user.
A pseudo algorithm of the loop is shown in Algorithm~\ref{alg:sync-loop}.

Each client has a centralized logical clock to track changes in its associated tasks~\cite{lamport:clock}.
Changes to a task also increment the clock of the corresponding client and the updated value is published over MQTT.
On the client device, whenever the sync loop notices that its local logical clock has fallen behind, it will request its state from the server, as shown in the first case-clause of Algorithm~\ref{alg:sync-loop}.
In response, the client receives its current logical clock and all active tasks.
Each task has an ID and the number of results submitted.

Three functions are not defined in Algorithm~\ref{alg:sync-loop}:
\begin{itemize}
    \item $fetchState$ requests the client state from the server.
    \item $submit$ uploads results or status changes to the server.
    \item $syncContainers$ starts and stops containers to match the currently active tasks.
\end{itemize}
Both $fetchState$ and $submit$ send a new state back to the sync loop.
However, only one of them is allowed to run at a time, which is controlled by the $syncingState$ boolean.
A dirty state can arise if new results or statuses are received from the result handler or a container supervisor (see Figure~\ref{fig:client}) while $syncingState$ is set to true.
When this happens, the newly received results or statuses are not visible to the active $submit$ thread.
Therefore, when the $dirtyState$ flag is true and the active $submit$ thread sends a new state to the sync loop, the algorithm calls $submit$ again.

\RestyleAlgo{ruled} 

\begin{algorithm}
\DontPrintSemicolon
\caption{The sync loop} 
\label{alg:sync-loop}
\SetKwBlock{Loop}{loop}{} 
\KwData{\\A state containing a logical clock $ts$, $tasks$ info about active tasks at $ts$, and a map from tasks to results and a status called $localTasks$.\\
Two booleans, $syncingState$ and $dirtyState$, initialized to false.}

\Loop{ 
\Switch{event}{
  \Case(\tcp*[f]{Notified about a change from a user}){received logical clock \textbf{tsR} from MQTT}{
    \If{$tsR > state.ts$}{
        $state.ts \gets tsR$\;
        \If{not syncingState}{
            $syncingState \gets true$\;
            Spawn fetchState(cloned state)\;
        }
    }
  }
  \Case{received new state \textbf{s}}{
    \eIf{$s.ts \geq state.ts$}{
        $state.ts \gets s.ts$\;
        $state.tasks \gets s.tasks$\;
        \eIf(\tcp*[f]{Not done; a task in $state.localTasks$ changed while syncing}){diryState}{
            $dirtyState \gets false$\;
            Spawn submit(clone(state))\;
        }(\tcp*[f]{We are done syncing state with server}){
            $syncingState \gets false$\;
            Spawn syncContainers(s)
        }
    }{
        Spawn fetchState(clone(state))\;
    }
  }
  \Case(\tcp*[f]{Only used by $syncContainers$}){received local tasks \textbf{L} from syncContainers}{
    $state.localTasks \gets L$\;
    $syncingLocals \gets false$\;
  }
  \Case(\tcp*[f]{Local changes}){received result \textbf{r} or status \textbf{s} from container for task \textbf{t}}{
    append $r$ or change to $s$ for $state.localTasks[t]$\;
    \eIf{syncingState}{
        $dirtyState \gets true$\;
    }{
        $syncingState \gets true$\;
        Spawn submit(clone(state))\;
    }
  }
}
}
\end{algorithm}

\subsubsection{Containerization}
A new Docker container is started when $syncContainers$ sees a new task not present in the $localTasks$ map.
The procedure to run and supervise the container is spawned in a new thread.
The supervisor thread sends an \lstinline{ERROR} status for the task back to the sync loop if the container exits with an error signal, otherwise, it sends a \lstinline{FINISHED} status.
Also, it can stop the container if the task has been canceled or removed.

\subsubsection{Result handler}
The payload running in the container can communicate with the supervising client through an API implemented as a gRPC server---the Result Handler as seen in Figure~\ref{fig:client}.
Depending on which RPC is used, results are either forwarded to the sync loop for publication to the server or stored locally as intermediate results to be loaded later for further processing.
Storing an intermediate result is a way of handling client restarts. 

\subsubsection{Signal handler}

Our initial hardware target is the Host Mobility MX-4 unit running the Wireless Information Collection Environment (WICE) platform~\cite{alkit:wice}.
The WICE platform enables remote deployment of our client binary and exposes a Signal Broker API that allows us to consume signals from the CAN and FlexRay buses through a publish-subscribe model.

As shown in Figure~\ref{fig:client}, the Signal Handler is an AutoSPADA client component that subscribes to signal values from the WICE Signal Broker. 
Internally, the Signal Handler consists of a gRPC server and a signal broker proxy to consume and cache signals from the external WICE Signal Broker.
This proxy allows us to normalize the behavior between different external signal sources.
That is, the API used by tasks to read signal values stays the same even if support for another signal source, e.g., MQTT, is added.

\section{Introduction to Using AutoSPADA}
The platform is ready for use after it is deployed on a cloud platform and to clients.
We provide two libraries for Python that wrap gRPC calls and let users consume message queues.
One library enables users to work interactively with the platform, while the other is used in payloads running on the client.
This section briefly demonstrates what users can express with the two libraries.

\subsection{The Python client library}

The client library provides functions used to define AutoSPADA payloads.
The core functionality lets users read signal values and publish results.
Additionally, one can read task parameters, cache a state as an intermediate result, and read a previously cached state.
State caching is local to the task and will be removed upon task completion, but crucially survives client restarts.
Listing~\ref{code:payload} shows a simple payload using the core functionality and two task parameters. 

\begin{lstlisting}[
float,floatplacement=H,
language=Python,
label=code:payload,
caption=A payload to compute the mean of value readings collected over a configurable period of time. Results are published as a JSON-serializable Python dictionary.
]
import autospada
import time

seconds_to_collect = autospada.parameters['seconds']
signal_name = autospada.parameters['signal_name']

total, count = 0, 0
start_time = time.monotonic()
while time.monotonic() - start_time < seconds_to_collect:
    total += autospada.next_signal(signal_name)
    count += 1

mean = total / count
autospada.publish({"Mean": mean, "n_values": count})
\end{lstlisting}

There is no limit on the number of results a task may publish.
This is important because it allows users to define long-running, potentially indefinite, tasks that publish results periodically.
In the case of long-running tasks, the need to cache an intermediate result or state becomes evident.
For example, consider a task that builds a histogram over time and periodically publishes its progress.
Without caching, the task state is lost when the vehicle is turned off, forcing the task to start over from the initial state when the vehicle starts up again.
Because the binning is reset, the user has to go through all published results to find the total bin counts over the lifetime of the task.
However, by also caching the histogram binning when publishing results, only the latest result is needed since the historical bin counts now increase monotonically.
This simplifies the off-board analysis and reduces the required data transfer from the cloud to the user.

\subsubsection{Testing payloads}
Writing payloads, or code in general, is an error-prone endeavor.
Therefore, testing your code is always a good idea.
With an interpreted language like Python, it is easy to directly run the code and see if there are any immediate problems.
Users of AutoSPADA can test their Python payloads locally, without access to any clients or even a server node, in two ways: run the script directly or load it into the same kind of container used by clients.

By default, the \lstinline{autospada} library acts as a dummy library that returns random values for any signal and prints messages to standard output when side effects occur.
Payloads can, therefore, run independently like any other Python script.
However, the logical flow of a payload may depend on observing specific signal values.
In those cases, reading random values for all signals can prevent the payload from exiting normally.
Nevertheless, this simple test can still catch unexpected behaviors such as syntax errors.

A more robust testing methodology is to run the payload in a container using the same Docker image as a client would.
The user library has functionality that makes this simple to run, provided that the Docker image is also available for the user's local CPU architecture.
Moreover, this approach allows users to control the values of signals by providing a CSV file with hard-coded signal values.
In this way, payloads can be tested against specific inputs.

\subsection{The Python user library}
As shown in Figure~\ref{fig:assignment-db}, users can create documents for assignments, payloads, parameters, and tasks.
The user library provides functions to retrieve and create these documents.
With a connection to the RabbitMQ server, the library also lets users subscribe to new results and changes in task statuses using either blocking await functions or lazy iterators.

\subsubsection{A simple example}
The following section breaks down a simple user workflow.
We demonstrate how to define an assignment that instructs online clients---in this case, vehicles---to run a task to compute their mean speed over five seconds. 

The Python library for the user is called \lstinline{autospada_user} and provides the \lstinline{User} class through which all actions to the server are made.
A \lstinline{User} object is initialized with a configuration file that contains the information required to connect and authenticate with the AutoSPADA cloud.
\begin{lstlisting}[language=Python]
import autospada_user
user = autospada_user.User('user_config.toml')
\end{lstlisting}

A payload object is created below with the contents of a Python file, similar to the one defined in Listing~\ref{code:payload}.
The library refers to payloads, parameters, tasks, and assignments as document objects, and this payload object has not yet been committed to the database. 
\begin{lstlisting}[language=Python, firstnumber=3]
from pathlib import Path
code = Path('mean_payload.py')
payload = user.payload(code.read_text(), name='Average')
\end{lstlisting}

Parameters are given as a separate object for composability.
\begin{lstlisting}[language=Python, firstnumber=6]
parameters = user.parameter(
    {"seconds": 5, "signal_name": can_speed_name})
\end{lstlisting}

Here, identical tasks are prepared for all clients that are currently online.
\begin{lstlisting}[language=Python, firstnumber=7]
clients = user.get_clients(online_only=True)
tasks = []
for client in clients:
    tasks.append(user.task(
        client.id, payload, parameters))
\end{lstlisting}

Since all tasks serve the same purpose for each vehicle, they are grouped into one assignment.
An assignment does not need to have any related tasks, but every task needs an assignment for the sake of consistency.
\begin{lstlisting}[language=Python, firstnumber=11]
assign = user.assignment("Mean speed", tasks)
\end{lstlisting}

The \lstinline{commit} method commits the assignment document object to the database, including all related documents if they have not been committed yet.
The library is designed with \textit{method chaining} in mind~\cite{method:chaining}, and the assignment object is therefore returned from \lstinline{commit} so that \lstinline{await_results} can be called directly.
This final line will wait for all tasks to finish and then return all results.
\begin{lstlisting}[language=Python, firstnumber=12]
results = assign.commit().await_results()
\end{lstlisting}

\section{Evaluation}

We have conducted an experiment to evaluate the overhead of AutoSPADA under idealized conditions.
The experiment aims to demonstrate that the system can be used interactively on relevant hardware.
If users have to wait for minutes for a simple task and its results to propagate through the system even in an idealized scenario, we can hardly claim that AutoSPADA enables users to develop interactively.

\subsection{Measurements}

The overhead is measured in terms of latency from the user's perspective.
The design of the experiment is such that we interpret the measurements as proxies for events in the system. 
This is achieved using two unconventional tasks---one that publishes empty results twice in succession and then exits, and another that does nothing.
None of the tasks specify parameters, which saves an additional communication round to the server.
Using these two tasks, we proceed to take four measurements that we will refer to as $t_{start}, t_{delay}, t_{exit}$, and $t_{cycle}$.

\subsubsection{Task startup}
The first measurement, $t_{start}$, is taken from the time that the task is committed to the time that the first result is received.
To minimize overhead, we use a payload that immediately publishes empty results.
The user subscribes to receive assignment results on an AMQP queue before issuing the task.
Hence, results from the server are received as fast as possible.
Because the task publishes an empty result as soon as it starts, we interpret this measurement as the time to start the task plus the time to propagate a result back to the user. 

The amount of network communication included in this measurement totals three gRPC calls (user to server and two calls from client to server), one MQTT message, and one AMQP message.
The measurement also includes the time to start a Docker container.

\subsubsection{Delay between results}
In an idealized scenario, the time between observing two results published immediately after one another, $t_{delay}$, should be small.
We expect this to show that the system is highly responsive once the task is running in its container.
Because this measures a difference between two publishing events, the amount of communication included in $t_{delay}$ is only the inter-process communication from the task container to the client program, plus any difference in propagating the result from the AutoSPADA client to the user.
Since this approximates the overhead of inter-process communication between the client and the task container, it should be considerably smaller compared to the other measurements.

\subsubsection{Container shutdown}
The time from receiving the second result to observing a \lstinline{FINISHED} status, $t_{exit}$, is also measured.
The resulting number approximates the time it takes to exit the task container.
This, combined with $t_{start}$, adds some context to $t_{cycle}$.
The communication overhead included in this measurement is between the Docker daemon and the AutoSPADA client to detect shutdown, plus any difference in propagating the status from the client to the user.

\subsubsection{Task cycle}
Lastly, we measure the whole task life cycle, $t_{cycle}$, from \lstinline{task.commit()} to a user-observed \lstinline{FINISHED} status.
The goal of this measurement is to see how fast the smallest possible task propagates through the system. 
Whereas the previous measurements were taken at different stages of the same task, $t_{cycle}$ uses another task with a payload that only imports the \lstinline{autospada} library and then exits.
The import is included to make the measurement comparable to $t_{start}$ since the import loading time is not negligible.
The amount of network communication included in this measurement is the same as for $t_{start}$.
The measurement additionally includes the time to start and stop a Docker container.

\subsection{Experiment setup}
A minimal deployment of one client, one server node, and a user participated in the experiment.
The relevant specifications of each node are given in Table~\ref{tab:experiment-setup}.
A Raspberry Pi acted as the client node to better represent the processing capabilities of hardware that we envision could run the AutoSPADA client while being a consumer product.
In particular, the Raspberry Pi 3 Model B has similar specifications to the Host Mobility MX-4 T30 used during the AutoSPADA project.

The AutoSPADA client is deployed as a binary with a configuration file and valid certificates.
The compiled AutoSPADA client binary for ARM is \qty{32}{\mebi\byte}.
The \lstinline{top} utility reported \qty{26.0}{\mebi\byte} resident (RES) and \qty{20.5}{\mebi\byte} shared (SHR) memory sizes for a newly started idling client program.
The Docker image used was \qty{48}{\mebi\byte} and contained Python 3.8 plus our client library. 
This image was downloaded to the local Docker image registry before the start of the experiment.

\begin{table}\centering
\begin{tabularx}{\linewidth}{@{}l X X@{}}\toprule
Node & Host & Specifications \\ \midrule
User & Local VM & Ubuntu 22.04 LTS, Ethernet internet connection \\
Server & E2 x86-instance in GKE & Limited to 1 vCPU and \qty{1}{\gibi\byte} memory \\ 
Client & Raspberry Pi 3 Model B Rev 1.2 & Broadcom BCM2837, \qty{1}{\giga\byte} RAM, Raspberry Pi OS (32-bit Bullseye), WiFi internet connection \\ 
\bottomrule
\end{tabularx}
\caption{Specifications of the experiment participants. The server node was provisioned through Google Kubernetes Engine (GKE).}
\label{tab:experiment-setup}
\end{table}

The experiment was repeated 100 times to give aggregated statistics.
The user starts by preparing payloads and assignments so that their submission time is not included in any measurement.
Before tasks are committed, the user also creates connections to message queues to which results and task statuses are published.

New payloads were created in each iteration since the client keeps a cache of the ones most recently used.
This means that if payloads are reused in the experiment, they do not have to be downloaded again.
Hence, caching improves the $t_{start}$ and $t_{cycle}$ measurements noticeably.
Although payload reuse is efficient, including the payload download in the measurement is more representative of an iterative development process where payloads frequently change.
Therefore, we include the time to download the payload in our measurements. 

\subsection{Results}

The experiment results are presented in Table~\ref{tab:experiment-results}.
The difference between $t_{start}$ and $t_{cycle}$ suggests that container shutdown invokes a delay in the order of seconds, which is confirmed by the measured $t_{exit}$.
The time from submitting a task to observing a result, $t_{start}$, is roughly between 4 to \qty{4.5}{\second}, which is justifiable considering that it includes the time for container creation and startup.
The delay between two results, $t_{delay}$, is stable during the experiment, which is expected because identical gRPC calls between a task container and the client program should be nearly constant in time.
All measurements besides $t_{delay}$ involve container creation, teardown, or both, which is less predictable than an inter-process gRPC call and is observed as larger variances.

The peak resident memory size during the execution of the experiment was approximately \qty{29.0}{\mebi\byte}---up \qty{3.0}{\mebi\byte} from the idle client.
This number was taken from the \textit{high water mark} field (VmHWM) in the \lstinline{status} file of the AutoSPADA client process in the \lstinline{/proc} filesystem after the experiment was completed.
Importantly, this only measures the AutoSPADA client program and does not account for the memory used by Docker and the containerized task, c.f. Figure~\ref{fig:structurizr-architecture}.

\begin{table}\centering
\begin{tabular}{@{}lcccc@{}} \toprule
& \multicolumn{4}{c}{Measurements (s); $n=100$} \\ 
\cmidrule{2-5}
& $t_{start}$ & $t_{delay}$ & $t_{exit}$ & $t_{cycle}$ \\ \midrule
Mean       &   4.282 &          0.261 &      1.198 &     5.640 \\
SD         &   0.260 &          0.080 &      0.316 &     0.377 \\
$P_{5\%}$      &   3.973 &          0.233 &      0.830 &     4.940 \\
$P_{95\%}$     &   4.605 &          0.271 &      1.695 &     6.191 \\
\bottomrule
\end{tabular}
\caption{Results of the latency experiment. The statistics are derived from one hundred data points per measurement. Measurement times are given in seconds, and SD is a shorthand for the sample standard deviation. We also show the 5th and 95th percentiles since the data contain a few outliers at both ends.}
\label{tab:experiment-results}
\end{table}

\section{Related Work}
Various edge computing platforms already exist on the market---some build on known architectures such as Kubernetes, some focus on stream processing, while others have more unique designs.
This section highlights a few relevant actors and reviews their similarities and differences compared to the AutoSPADA platform.
A summary is given in Table~\ref{tab:platforms}.

\begin{table*}[!ht]
\centering
 \begin{tabular}{@{}l c c c c c c@{}} \toprule
   & AutoSPADA& IoFog & Stream Analyze & KubeEdge & Azure IoT Edge & AWS IoT Greengrass\\ [0.5ex] 
 \midrule 
 Open source & - & Yes & No & Yes & Partially & Partially\\ 
 Containerized workloads & Yes & Yes & No & Yes & Yes & Possibly \\
 Memory (\unit{\mega\byte}) & 27.3 & 256 & 0.017-5 &  40 & - &  96 \\
 Implementation language & Go & Java & C & Go & C\# & Java \\
 \bottomrule
 \end{tabular}
 \caption{A comparison of properties for different edge computing platforms. Memory footprints are either measured (KubeEdge~\cite{cilic2023:tool-eval} and AutoSPADA) or taken from the specified system requirements of the respective platform.}
 \label{tab:platforms}
\end{table*}

\subsection{Stream Analyze}

The Stream Analyze Platform from Stream Analyze Sweden AB is a commercial platform for collecting and aggregating data from edge devices, with a focus on stream processing~\cite{streamanalyze}.
Analytical models are sent to clients that execute the requests and send streaming results to a backend that users can query.
Data analysis is performed in the Object Stream Query Language (OSQL), their proprietary language for data stream processing, or using a graphical tool for query editing and result visualization.
In addition to streaming, Stream Analyze also supports offline retrieval of task results.

\subsection{KubeEdge}

Initially proposed by researchers at Huawei in 2018~\cite{xiong2018:kubeedge}, KubeEdge is an open-source edge computing framework built on Kubernetes.
The goal of the framework is to extend Kubernetes clouds to also include edge hosts, meaning that existing Kubernetes workloads can be applied to both cloud and edge workers.
The framework was accepted as a Cloud Native Computing Foundation (CNCF) incubating project in September 2020.

On a high level, KubeEdge has two components: CloudCore and EdgeCore.
The CloudCore is a centralized component that orchestrates edge devices.
The EdgeCore consists of several components that synchronize device status with the cloud, enable containerized execution of workloads, and mappers to allow external communication with the EdgeCore over common IoT protocols.
The EdgeCore connects to the cloud through a WebSocket client and has local storage to let the EdgeCore function when offline.

\subsection{Eclipse IoFog}

Eclipse IoFog is an open-source edge computing platform initially developed by Edgeworx and later donated to the Eclipse Foundation~\cite {Desbiens2023:edgecomputing}.
Edge devices run an IoFog component called Agent---a daemon service responsible for running containerized microservices.
Devices running the Agent are orchestrated by a component called Controller that constitutes the platform's control plane.
The Controller can be placed anywhere, even on the same device as an Agent, as long as it is reachable by all devices running the Agent.
Moreover, the control plane can be deployed to a Kubernetes cluster where the number of Controller instances can be scaled for high availability.

The components of IoFog are Java programs, meaning that they run on the Java Virtual Machine rather than natively.
Running a virtual machine generally imposes a performance penalty, especially in terms of memory. 
This is also observed in benchmarks where virtual machine languages used $2.28$ times more memory on average compared to natively compiled languages~\cite{Pereira2017:energy}.
Another evaluation showed that the IoFog Agent component uses approximately \qty{240}{\mega\byte}, nearly five times more memory than the corresponding component in KubeEdge and K3s~\cite{cilic2023:tool-eval}. 
This memory usage may be prohibitively large.

\subsection{Big Tech platforms}

Of the so-called Big Tech companies, three (Alphabet, Amazon, and Microsoft) offer cloud computing services with various edge or IoT-oriented products.
Amazon's AWS IoT Greengrass and Microsoft's Azure IoT Edge are edge computing platforms with open-source software components.
Their client runtimes are open source, but cloud integration relies on the respective provider's commercial services. 
Having retired the Google Cloud IoT Core, Alphabet now offers Google Distributed Cloud Edge (GDC Edge) instead.
However, GDC Edge is a fully managed service, meaning that Alphabet supplies both hardware and software, making it less relevant to include in our survey.

The Azure IoT Edge and AWS IoT Greengrass platforms are in many aspects similar to each other.
Both share the goal of bringing computation closer to data sources and support the deployment of Docker containers to edge devices.
Also, both platforms extend the respective cloud provider's services, such as AWS Lambda Functions.
The software running on edge devices is packaged into AWS IoT Greengrass Components or Azure IoT Edge Modules, respectively.
Amazon provides a library of prebuilt Greengrass Components and Microsoft similarly provides prebuilt IoT Edge Modules such as Azure Stream Analytics.
Microsoft provides Software Development Kits (SDKs) for users to write custom IoT Edge modules. 
Likewise, Amazon provides SDKs to allow users to develop custom Greengrass Components.
A difference between the two is that IoT Edge Modules run as containers, whereas AWS Greengrass Components are not containers by default.

\subsection{Comparison}

AutoSPADA distinguishes itself from other cloud or Kubernetes-based platforms because of its focus on interactive use.
In particular, AutoSPADA users work entirely in Python, without the need to write Kubernetes manifests or build Docker images for foreign CPU architectures.
This simplifies the development cycle, encouraging exploration and rapid prototyping.

AutoSPADA and Stream Analyze share a focus on computational tasks rather than microservices. 
As a result, they are not designed to deploy general services such as API servers or databases, and instead run their computational tasks in isolation.
For example, AutoSPADA has no network path between tasks, and users cannot directly access running task containers.
In contrast, microservice architectures, such as KubeEdge or ioFog, require device-to-device communication because services must be able to reach each other even if they run on separate devices.
In general, microservice orchestrators view connected nodes as processors to which services can be freely scheduled.
Likewise, client nodes are agnostic to the services they run.
However, in edge computing, where data comes from a multitude of heterogeneous sources, prior knowledge of which devices a workload can be scheduled for (e.g., that the device has the expected sensors) is necessary.
No matter the platform, this information has to be known or sourced before creating the task or service.

Among the surveyed platforms, Stream Analyze requires the least amount of client resources by a wide margin.
The very small memory footprint of its client runtime allows it to run in extremely resource-constrained environments.
The runtime can also exist without an operating system in a so-called bare-metal deployment.
The other platforms, AutoSPADA included, need sufficient client resources to run a Linux environment to containerize their workloads.

\section{Discussion and Future Work}
The AutoSPADA platform has been deployed in various stages during its development process.
The client has been deployed to the cloud, to standalone hardware, and into the same hardware in cars through the WICE binary deployment.
We also held a final demonstration and workshop, where we remotely deployed the platform onto live vehicles and guided our collaborators from Volvo Cars to use it.
During the workshop session, we explored how our collaborators could express their use cases with AutoSPADA.
Two use cases were considered: one related to durability analysis and another to measuring the impact of one-pedal drive on energy consumption in an electric car. 
On the European Union's version of the technology readiness level (TRL), the platform is considered a level 7 because it was demonstrated running on vehicles~\cite{trl:eu}.
A website from the European Commission also exemplifies TRL 7 for system technologies as: "Testing is moved to operational environments such as a vehicle or machines"~\cite{euraxess:trl}.

Throughout the project, we implemented and tested various use cases for testing and demonstration purposes.
Two of them were \textit{active learning} and \textit{mapping hazard spots}.
Active learning is a machine learning field where the training model itself is used to decide what data is most informative to collect for labelling~\cite{settles2009:AL}.
This case demonstrated how machine learning inference could be used on edge devices.
In the mapping hazard spots case, we updated an external database to create a map of slippery roads.
Although this particular type of service already exists, e.g., the Road Surface Alerts by NIRA Dynamics~\cite{nira:slippery}, the mapping hazard spots use case demonstrated the efficiency and simplicity of interactively prototyping fleet-wide services using AutoSPADA.

Rapid prototyping in AutoSPADA is facilitated by the provided Python libraries that are designed to be intuitive for anyone with Python experience.
Users can interact with AutoSPADA entirely through Python, allowing them to work conveniently in, e.g., interactive Jupyter notebooks~\cite{marijan2021:jupyter}.
Our experiment shows that the latencies in the system are sufficiently low to suggest that the platform is suitable for interactive use.
Moreover, our client program performs well compared to other platforms on the market.

\subsection{Future considerations}
Other languages may offer better performance or compatibility for running any of the services on the platform.
Since the interfaces between the user, server, and client nodes are specified using protobuf, they can easily be compiled to several other target languages.
This allows for a switch of implementation language should it ever be needed.

Several features to enhance the AutoSPADA platform could not be implemented within the duration of the AutoSPADA project. 
We conclude by highlighting some valuable additions that could be made to the platform in the future.

\subsubsection{Data abstraction}

Because our partners in the project in which the AutoSPADA platform was developed are in the automotive industry, our focus has been on connected vehicles.
Specifically, reading device data through the \lstinline{get_signal} function, as shown in Listing~\ref{code:payload}, only works if the clients are WICE-equipped vehicles.
For other hardware platforms, users must include code to read the desired signals in their payloads. 
Making the data subscription service configurable to support more protocols, e.g., MQTT would enable the use of this function on a wider range of client devices.

\subsubsection{Resource management}

The AutoSPADA platform cannot manage clients through the user API.
For example, you cannot prune Docker images on clients through the current user API. 
Redundant Docker artifacts should be removed regularly since client devices are resource-constrained.

Because tasks run in Docker containers, each task's resource usage can be limited. 
The maximum amount of resources used per task should be a configurable parameter on the assignment level. 
Moreover, AutoSPADA currently does not assess the busyness of clients, which means that clients are vulnerable to simple denial-of-service attacks from malicious or simply inattentive users. 
A mechanism to refuse or schedule pending tasks for clients with exceeded resource quotas should be implemented.

\subsubsection{Customizable container environment}

Only the Python standard library and the AutoSPADA library are available for use in payloads.
However, users surely want to leverage Python's rich ecosystem of third-party packages when writing payload code.
This means that the Docker image used to containerize tasks on clients needs to be customizable in some way.
Though it is already possible to change the image, it is impractical since it requires a manual reconfiguration and redeployment of the AutoSPADA client on each edge device.
A more flexible approach is to configure the Docker image to use for task containerization per assignment.

Allowing custom Docker images also raises the question of how package customization can be implemented without increasing the complexity too much for users.
For example, the threshold for non-expert users to build multi-platform Docker images of their own is high.
Instead, one could have up to three tiers with increasing flexibility and complexity.
In the first tier, we provide pre-built images with different sets of common packages, such as the ubiquitous NumPy package for array programming~\cite{harris2020:numpy}.
A second tier would allow users to specify their required dependencies via, e.g., a \lstinline{pyproject.toml} file and then build the image centrally.
This adds server-side complexity and cost since a central build server is required.
Finally, we could allow users to build completely custom Docker images using the AutoSPADA base image.
However, this adds user-side complexity, especially when images have to be built for a different CPU architecture.
Also, related to device resource management, placing custom Docker images on edge devices makes image pruning an essential feature.

\subsubsection{Client metadata}
Users are fully responsible for assigning tasks to clients and making sure that the client is equipped with the expected hardware and sensors.
This typically means that a user queries the platform for all online clients and cross-references that list with external information about each client device.
However, it would be convenient for users if they only need to specify client requirements instead of forcing task placement onto a specific device.
After all, the user might only require that a specific sensor exists.
Also, if the server can decide on task placement, tasks can be held in a pending state until a compatible client becomes available. 
This would avoid forcing users to guess which client to choose if all compatible clients are offline. 
Besides the additional server-side logic, such a mechanism requires clients to register metadata about themselves, such as the device type, CPU architecture, and available sensors.

\section*{Acknowledgements}
This research was financially supported by the project Automotive Stream Processing and Distributed Analytics (AutoSPADA) in the funding program FFI: Strategic Vehicle Research and Innovation (DNR 2019-05884), which is administered by VINNOVA, the Swedish Government Agency for Innovation Systems.

We are grateful to AI Sweden for providing remote access to MX-4 units through the Edge Learning Lab. The Edge Learning Lab has been a valuable test environment during the development of the AutoSPADA platform.

Special thanks to Anders Nord and Viktor Larsson at Volvo Cars for contributing the durability and one-pedal drive use cases, respectively, and for their collaboration in testing the platform on Volvo cars.

\section*{Declaration of Generative AI and AI-assisted technologies in the writing process}
During the preparation of this work, the authors used GrammarlyGO and Bing Chat in order to quickly improve the readability and flow of the text. After using either tool, the authors reviewed and edited the content as needed and take full responsibility for the content of the publication.




\bibliographystyle{elsarticle-num} 
\bibliography{references}


\end{document}